\documentclass[aps,prd,onecolumn,groupedaddress,nofootinbib,superscriptaddress,preprintnumbers]{revtex4}  
\usepackage{graphicx,mathtools,tabu}
\usepackage{epstopdf}
\usepackage{amsmath}
\usepackage{amsfonts}
\usepackage[colorlinks=true,citecolor=red,linkcolor=blue,breaklinks=true]{hyperref}
\usepackage{accents}
\usepackage[figure, table]{hypcap}
\usepackage{amssymb}
\usepackage{appendix}
\usepackage{comment}
\usepackage{bbold}
\usepackage{color}
\usepackage{slashed}
\usepackage{subfigure}
\usepackage{setspace}
\usepackage{footnote}
\usepackage{multirow}
\usepackage{braket}
\usepackage[normalem]{ulem}
\usepackage[a4paper, portrait, margin=0.8in]{geometry}
\usepackage[utf8]{inputenc}
\usepackage{mathrsfs}
\usepackage{longtable}
\usepackage{enumitem}

\LTcapwidth=\textwidth

\usepackage{url}
\usepackage{wrapfig}
\usepackage{braket}

 \def\be   {\begin{equation}}  
  \def\ee   {\end{equation}}
 \def\ba   {\begin{array}}     
  \def\ea   {\end{array}}
 \def\bea  {\begin{eqnarray}}  
  \def\eea  {\end{eqnarray}}
 \def\bean {\begin{eqnarray*}}  
 \def\eean {\end{eqnarray*}}

  \def\be {\beta}

\begin{document}

\title{Reactor Antineutrino Oscillations at Super-Kamiokande}

\author{Andr\'{e} de Gouv\^{e}a}
\email{degouvea@northwestern.edu}
\affiliation{Northwestern University, Department of Physics \& Astronomy, 2145 Sheridan Road, Evanston, IL 60208, USA}
\author{Ivan Martinez-Soler}
\email{ivan.martinezsoler@northwestern.edu}
\affiliation{Northwestern University, Department of Physics \& Astronomy, 2145 Sheridan Road, Evanston, IL 60208, USA}
\affiliation{Theory Department, Fermi National Accelerator Laboratory, P.O. Box 500, Batavia, IL 60510, USA}
\affiliation{Colegio de F\'isica Fundamental e Interdisciplinaria de las Am\'ericas (COFI), 254 Norzagaray street, San Juan, Puerto Rico 00901.}

\begin{abstract}
SuperKamiokande (SK) doped with gadolinium has the capability to efficiently identify electron-antineutrinos through inverse beta-decay. Given the size of SK and the number of nuclear reactors in its vicinity, we argue that SK can observe the oscillations of reactor antineutrinos driven by the so-called solar mass-squared difference $\Delta m^2_{21}$. After only one year of data taking, we estimate that SK can measure $\Delta m^2_{21}$ with enough precision to help inform the current small tension between existing results from KamLAND and solar neutrino experiments.
\end{abstract}

\preprint{NUHEP-TH/20-03, FERMILAB-PUB-20-222-T}

\maketitle

Our understanding of neutrino oscillations continues to evolve rapidly. Five out of the six oscillation parameters -- $\sin^2\theta_{12}, \sin^2\theta_{13},\sin^2\theta_{23}, \Delta m^2_{21}$, and $\Delta m^2_{31}$, following the  PDG parameterization \cite{Tanabashi:2018oca} -- are known at the few to several percent level~(see, for example, \cite{Esteban:2018azc}) and there is growing nontrivial information on the sixth parameter -- the CP-odd phase $\delta$ -- and the neutrino-mass ordering, which are among the main targets of the current and next generation of long-baseline oscillation experiments. Different experiments, with different sources and detector technologies, provide information on different subsets of oscillation parameters. $\Delta m^2_{31}$, for example, is constrained by measurements of neutrinos produced in accelerators, the atmosphere, and nuclear reactors, and make use of baselines that range from 1~km to 10,000~km. The fact that these different experiments point to the same range of values for $\Delta m^2_{31}$, along with the mixing parameters $\sin^2\theta_{13}$ and $\sin^2\theta_{23}$, serves as evidence that our current understanding of the neutrino sector is at least an excellent approximation of nature. 

As far as the so-called solar parameters -- $\sin^2\theta_{12}, \Delta m^2_{21}$ -- are concerned, the current situation is distinct. Since, it turns out, $\Delta m^2_{21}\ll|\Delta m^2_{31}|$, experiments sensitive to $\Delta m^2_{21}$ and $\sin^2\theta_{12}$ require very long baselines and very intense electron-(anti)neutrino sources. For all practical purposes, all information on the solar parameters comes from experiments that measure the solar neutrino flux and from the very-long-baseline reactor antineutrino experiment KamLAND. When it comes to measuring the solar parameters, the two different experimental efforts share the load more or less equally. KamLAND measures $\Delta m^2_{21}$ with higher precision \cite{Gando:2013nba} while the solar data~\cite{Abe:2016nxk,Cleveland:1998nv,Kaether:2010ag,Abdurashitov:2009tn,Aharmim:2011vm,Bellini:2014uqa,Bellini:2011rx,Bellini:2008mr} are more sensitive when it comes to measuring $\sin^2\theta_{12}$. The lack of more, different information on the solar parameters is the source of modest concern, especially given that the measurements of $\Delta m^2_{21}$ at KamLAND and solar experiments are under slight tension, discrepant at, roughly, the two-sigma level. The values of the solar parameters also play a role when it comes to extracting the value of the CP-odd phase $\delta$ from data from accelerator-based and atmospheric long-baseline experiments. In summary, more information on the solar parameters would be most welcome. 

The proposed Jiangmen Underground Neutrino Observatory (JUNO), a reactor antineutrino experiment with a baseline around 50~km, aims at measuring both $\sin^2\theta_{12}$ and  $\Delta m^2_{21}$ with sub-percent precision~\cite{An:2015jdp}. JUNO is currently under construction and plans to start collecting data in a couple of years. The next-generation experiments HyperKamiokande and DUNE will also provide new information on solar neutrinos~\cite{Capozzi:2018dat,Abi:2020evt,Abe:2018uyc}. It is widely anticipated that, by 2030, our understanding of the solar parameters will be quantitatively more precise and, perhaps, qualitatively distinct. 

In this note, we argue that the Super-Kamiokande experiment (SK), thanks to its recent upgrade, can significantly contribute to our understanding of solar neutrinos potentially before the end 2021. We estimate how well SK can measure the flux of reactor antineutrinos in Japan after gadolinium is added to the ultra-pure water and find that a nontrivial measurement of $\Delta m^2_{21}$ can be extracted from one year of reactor data. This is more than incremental information given that, while both KamLAND and the measurement proposed here use reactor antineutrinos, the detector technologies are different -- liquid scintillator versus gadolinium-loaded ultra-pure water -- and the distribution of nuclear reactors in Japan has changed significantly since the terrible March 2011 earthquake/tsunami event off the northeast of Japan. The idea of using water Cherenkov detectors doped with gadolinium to study reactor antineutrinos and measure the solar oscillation parameters was first proposed by the authors of Ref.~\cite{Choubey:2004bf}. 

At SK, the main sources of $\bar{\nu}_{e}$ with energies below $10$~MeV are several nuclear reactors located around the experiment~\cite{Abe:2018uyc}. Here, we consider all nuclear reactors that are less than 2,000~km away from SK as the antineutrino sources~\cite{Baldoncini:2014vda}. This includes nuclear reactors in
Japan, South Korea, and a few sources in China. We estimate that the flux of antineutrinos at SK from reactors located further away is negligible. 

In the last few years, some of the Japanese nuclear power plants that shut down after 2011 have restarted. In order to reliably estimate the reactor antineutrino flux arriving at SK today, we follow Ref.~\cite{Baldoncini:2014vda}. We consider all the reactor cores active in 2018, including the times when these were off. It turns out that, today, the nuclear reactors closest to SK are around 180~km away. This is different from the circumstances of most KamLAND data, dominated by nuclear reactors under 100~km away.\footnote{SK and KamLAND are, given the baselines of interest here, next to each another.} Hence, compared to the bulk of KamLAND data taking, the instantaneous reactor flux today is smaller at the SK site and the average baseline is longer. The smaller flux at SK today is partially compensated by a larger number of targets, roughly $40$ times larger than that at KamLAND.

For each reactor, the flux of $\bar{\nu}_{e}$ originates mainly from the beta-decays of the fragments of the fission-decay-chain of four isotopes: $^{235}$U, $^{239}$Pu, $^{238}$U and $^{241}$Pu. While the fractional contribution of each isotope depends on the reactor type, the main contribution comes from the first two isotopes, which account for almost $90\%$ of the neutrino production. When simulating the reactor neutrino flux,  we use the exponential-polynomial parametrization  provided in Refs.~\cite{Mueller:2011nm,Baldoncini:2014vda}. For each reactor type, we consider the charateristic fission-fraction associated to each relevant isotope~\cite{Baldoncini:2014vda}, assuming that their values remain constant during the time of  data taking. The impact of the time dependency of the flux due to the fuel composition of the nuclear reactor is small compared to other uncertainties in the flux so we expect this approximation to have negligible impact on the sensitivity to neutrino oscillatilons.

Super-Kamiokande is a 50~kton (22.5~kton fiducial mass) water Cherenkov detector that has been taking data for more than 20 years. 
It detects neutrinos by measuring the Cherenkov radiation emitted by charged particles created in the neutrino interactions.
It is currently undergoing an upgrade that will allow it to detect relatively small fluxes of electron antineutrinos with ${\cal O}(10~\rm MeV)$ of energy on top of radioactive backgrounds. In a nutshell, the idea is to dissolve a gadolinium salt in the ultra-pure water. Gd has a very large cross section for capturing neutrons and allows one to tag them very efficiently. Under these circumstances, low-energy antineutrinos can be detected via inverse beta-decay --  $\bar{\nu}_{e} + p \rightarrow n + e^{+}$ -- off of the protons that are readily available in H$_2$O. The positron can be detected (and its energy measured) along with a time-delayed neutron-capture. The positron--delayed-neutron coincidence allows one to veto the backgrounds that plague the detector at these energies. The addition of gadolinium to SK was motivated by the diffuse supernova neutrino background and estimates that its discovery is within reach of SK if one can effectively identify ${\cal O}(10~\rm MeV)$ electron antineutrinos \cite{Beacom:2003nk}. It took several years to determine how to safely add Gd to SK. 

SK loaded with Gd is also sensitive to reactor antineutrinos. For energies higher than several MeV~\cite{Abe:2016nxk}, the Cherenkov radiation emitted by the positron from inverse beta-decay is observable in SK. After the absorption of the neutron by a Gd nucleus, a prompt cascade of photons with energies around $8$~MeV -- visible in SK -- is emitted. The correlation in position (around 50~cm) and time (around $30\mu\text{sec}$)~\cite{Abe:2018uyc} between the positron and the photon cascade allows SK to positively identify inverse beta-decay events. The collaboration is currently doping SK with  $0.2\%$ of Gd, which will provide around $90\%$ efficiency for neutron capture~\cite{Ikeda:2015}.

In order to estimate the number of reactor antineutrino events, we assume that the neutrinos are observable if they interact inside the SK fiducial volume. The differential flux of electron antineutrinos at the detector is 
\begin{equation}
 \phi_{\bar{\nu}_{e}}(E_{\nu}) = \sum_{i}\phi_{i}(E_{\nu},L_{i})P_{ee}(E_{\nu},L_{i})\,,
\end{equation}
where $\phi_{i}(E_{\nu},L_i)$ is the differential antineutrino flux from the $i$-th nuclear reactor, located a distance $L_i$ from SK, and the sum is over all the reactors included in the analysis. $E_{\nu}$ is the neutrino energy and $P_{ee}$ is the electron antineutrino survival probability.

Keeping in mind that (a) reactor neutrino energies fall in the $1-10$~MeV range, (b) $L_i\gtrsim200$~km, (c) we are averaging over several nuclear reactors, and (d) SK has a finite energy resolution, discussed below, $\Delta m^2_{31}$-effects safely average out and, in vacuum, 
\begin{equation}\label{eq:osc}
  P_{ee}(E_{\nu},L) = c^{4}_{13}\left(1 - \sin^22\theta_{12}\sin^2\left(\frac{\Delta m^2_{21}L}{4E_{\nu}}\right)\right) + s^4_{13}\,,
\end{equation}
where $c_{13} = \cos\theta_{13}$ and $s_{13} = \sin\theta_{13}$, is an excellent approximation for the electron-antineutrino survival probabilty. Given what is known about $\Delta m^2_{21}$ and the energies of interest,  matter effects are very small. Nonetheless, since  the antineutrinos do propagate through the Earth, we include matter effects in our numerical analyses, making use of the Preliminary Reference Earth Model~\cite{Dziewonski:1981xy} for the matter density profile. 

We organize the simulated data in bins of the reconstructed prompt energy $E_p$, related to the neutrino and positron energies by $E_{p} = T_{e} + 1.022~\text{MeV} = E_{\overline{\nu}_{e}} - 0.78~\text{MeV} - T_{n}$~\cite{An:2016srz}, where $T_{e}$ and $T_{n}$ are the kinetic energies of the positron and the neutron, respectively, and $1.022$~MeV is the energy the pair of photons created when the positron annihilates with an electron in the medium inherits from the electron plus positron masses. We consider 16~bins of $0.5$~MeV each. The number of events in the $i$-th bin is 
\begin{equation}
N_{i} = N_{t}\int
dE_{\nu}dE_{e}\phi_{\bar{\nu}_{e}}(E_{\nu})\frac{d\sigma_{\text{IBD}}(E_{\nu},E_{e})}{dE_{e}}\epsilon(E_{e})M(E_{e},E_{p})\,,
\end{equation}
where $N_{t}$ is the number of targets and $\sigma_{\text{IBD}}(E_{\nu},E_{e})$ is the the cross section for inverse beta-decay, from Ref.~\cite{Strumia:2003zx}, as a function of the netrino energy $E_{\nu}$ and the positron energy $E_e=m_e+T_e$. We include antineutrinos with energies above $1.5$~MeV. We assume that for $E_e$ values above 4~MeV the detector sees the inverse beta-decay event with 100\% efficiency. Below this threshold, the detector efficiency $\epsilon(E_{e})$ is smaller; we use the results presented in Ref.~\cite{Simpson:2019xwo}.  The positron energy is related to the prompt energy via the migration matrix $M(E_{e},E_{p})$, computed via Montecarlo integration. In this computation, we assume that  the SK energy resolution ($\sigma_{E}$) is the same as the one for the  detection of solar neutrinos, $\sigma_{E_{e}}=-0.0839 + 0.349\sqrt(E_{e}) + 0.0397 E_{e}$~\cite{Abe:2016nxk}. For a positron with $E_{e}=4$~MeV, the uncertainty is of order $20\%$.

\begin{figure}[ht!]
\includegraphics[width=0.45\textwidth]{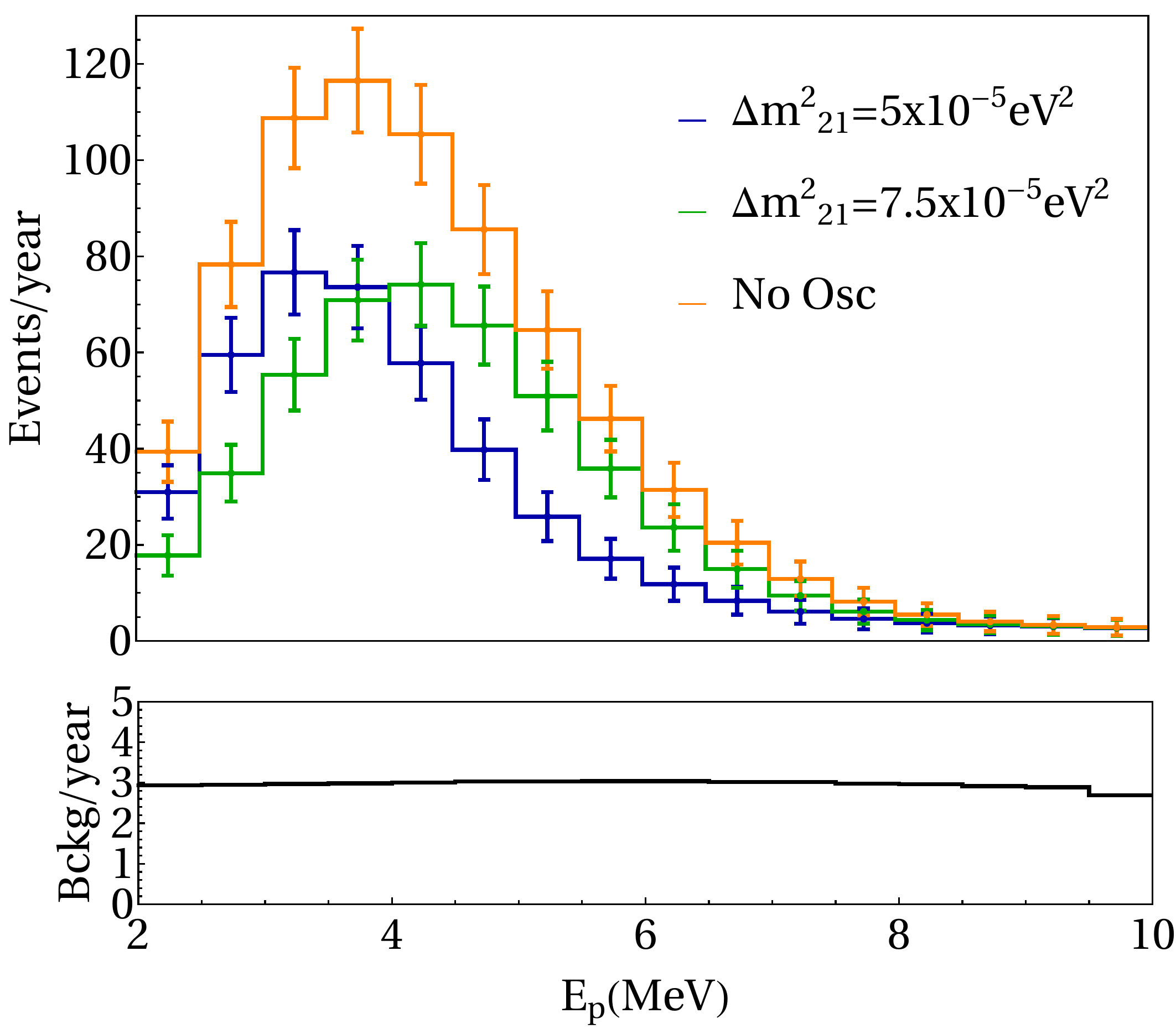}\,\includegraphics[width=0.45\textwidth]{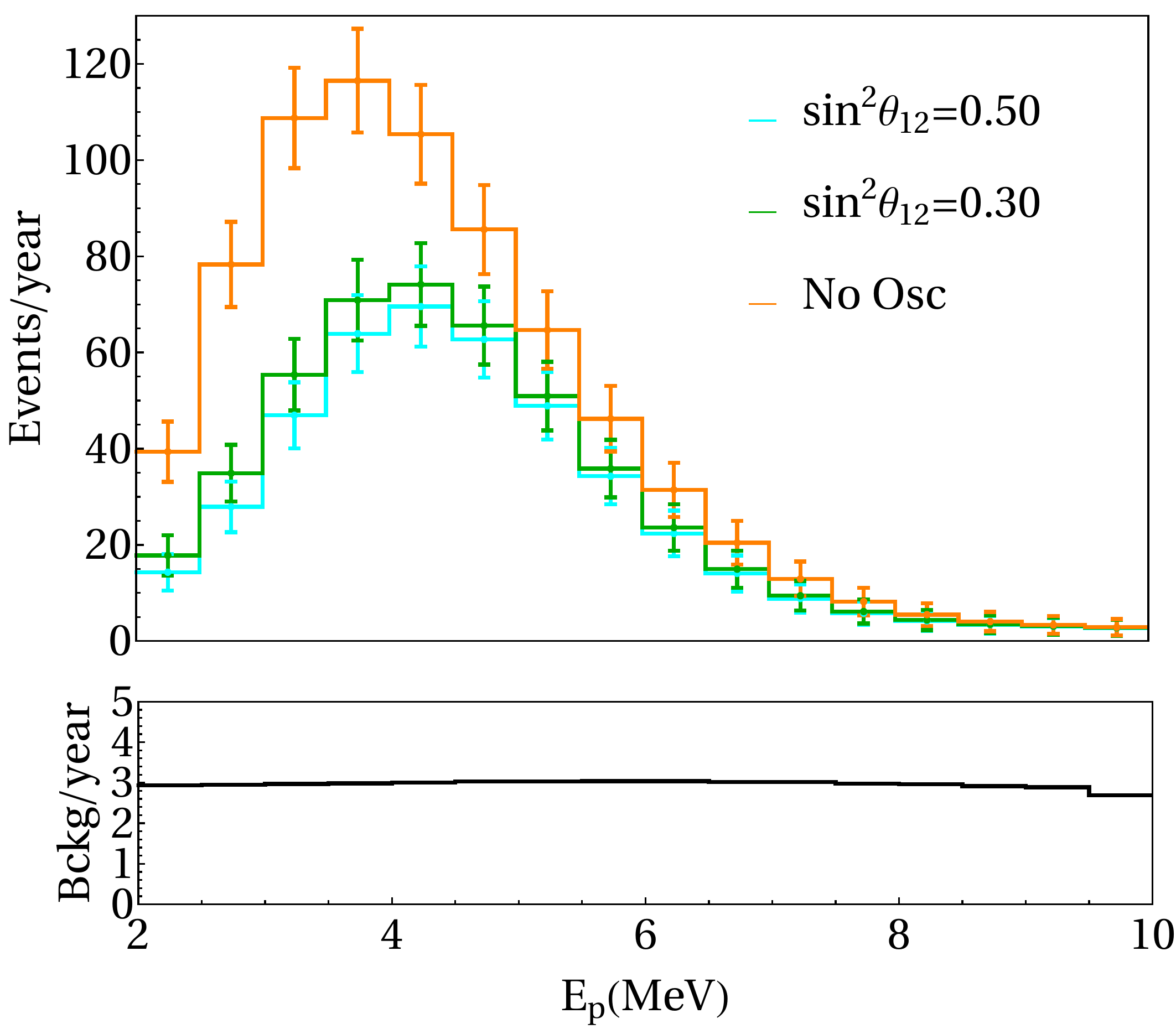}
\caption{Top: Simulated number of events for one year of data taking for two different values for the solar mass-squared difference, $\Delta m^2_{21}$, on the left, and two different values of the solar mixing angle, $\sin^2\theta_{12}$, on the right. The parameters that are not specified in each figure are fixed to their best-fit values from~\cite{Esteban:2018azc}, except for the panel on the right, where $\Delta m^2_{21}=7.5\times 10^{-5}\text{eV}^2$ is assumed. In yellow is the simulated number of events in the absence of oscillations. Bottom: Simulated number of background events for one year of data taking. See text for details.}
\label{fig:event}
\end{figure}

Figure~\ref{fig:event}(top) depicts the simulated number of events -- signal and background -- after one year of data taking for different values of the mixing parameters: (left) $\Delta m^2_{21} = (5\times 10^{-5}\text{eV}^2, 7.5\times 10^{-5}\text{eV}^2)$ and (right) $\sin^2\theta_{12} = (0.15, 0.30)$. In all plots, the other mixing parameters are fixed to their current best-fit values~\cite{Esteban:2018azc} except in Fig.~\ref{fig:event}(right), where we assume $\Delta m^2_{21}=7.5\times 10^{-5}\text{eV}^2$. The figure also depicts the spectrum one would encounter in the absence of neutrino oscillations. Thanks to the large values of $\sin^2\theta_{12}$ and the very long baselines, oscillation effects are readily observable after one year of data taking.  The shape of the spectrum -- e.g., position of the peak -- also changes once oscillation effects are included, indicating sensitivity to the value of $\Delta m^2_{21}$.

There are several sources of background~\cite{FernandezMenendez:2017cex,Simpson:2019xwo} that will impact the sensitivity to the oscillation parameters. Here we consider the following, which assume are dominant:
\begin{itemize}
\item Muon spallation. Cosmic-ray muons that find their way to the detector will produce radioactive isotopes via spallation. For example, interactions with $^{16}$O lead to a cascade that mimics the neutron-capture process, leading to false coincident events. This background can be reduced by taking advantage of timing and position information regarding muons in the detector, which can be used to veto muon-spallation-related events~\cite{Super-Kamiokande:2015xra}. Long-lived spallation products are of special concern since they are harder to veto efficiently.  This is the case of, for example, $^{9}$Li, which will undergo beta decay and emit neutrons. Techniques based on the energy profile in the isotope  production~\cite{Bays:2011si,Li:2014sea,Li:2015kpa,Li:2015lxa} have been proposed to reduce this type of background. From the expected contribution to HyperKamiokande~\cite{Abe:2018uyc} and taking into account the extra suppression of cosmic-ray muons due to the deeper location of SK, we estimate four events per year associated to this background source. In order to account for the uncertainties on the atmospheric muon flux, we include a Gaussian prior of $20\%$ in the normalization of this background in our simulated-data analysis.
  
\item Spontaneous fission from radioactive contaminants associated to the gadolinium doping. This signal mimics the antineutrino interaction with the production of photons with several MeV of energy and neutrons. The main contaminant is $^{238}$U. This background source can be reduced by controling the amount of uranium in the Gd-salt~\cite{FernandezMenendez:2017cex,Sekiya:2020pun}. Here we assume around ten events per year from the decay of $^{238}$U. Radon is another well known contaminant and has been studied exhaustively in the context of the detection of solar neutrinos~\cite{Abe:2010hy,Cravens:2008aa,Hosaka:2005um,Nakano:2019bnr}. Radon-related backgrounds are localized mainly outside of the fiducial volume and its impact can be mitigated with the purification system. The overall rate assumed here is 1\% of the standard radioactivity level of the Gd-salt, a reduction which can be obtained with the action of the resin AJ4400, discussed in Ref.~\cite{1352414}. This background level is potentially conservative. Dedicated studies~\cite{Ikeda:2019pcm} indicate that further significant reduction on this background is within reach. For this reason, we have not included an uncertainty related to estimate of this background in our simulated-data analysis.
  
\item Accidental coincidence of solar-neutrino interactions and radioactive beta-decays or neutron-production. The event rate for the coincidence with neutrons is estimated to be $8\times 10^{-6}/\text{day}$. Beta-decay processes can be misidentified as the absorption of a neutron by Gd. Here we estimate an accidental coincident rate of $ 33\, \text{events/year}$~\cite{FernandezMenendez:2017cex}. We considered the uncertainty in the flux of solar neutrinos~\cite{Abe:2016nxk} and found a negligible impact on our final results. For this reason, we have not included the uncertainty in the solar neutrino flux on the simulated data-analysis.  
 
\end{itemize}

Figure~\ref{fig:event}(bottom) depicts the simulated number of background events per year. We assume the shape of the energy distribution of background events related to spontaneous fission and coincidence of solar neutrinos with radioactive processes to be flat. This is a good approximation in the case of the interaction of solar neutrinos given the energy distribution of the flux at high energies. For the muon-spallation background, we make use of the background energy-distribution estimated for HyperKamiokande in Ref.~\cite{Abe:2018uyc}.

We analyze the simulated data using a Gaussian likelihood method. In addition to the uncertainty in the energy reconstruction, already discussed, we include a 4\%~\cite{Gando:2013nba} overall normalization uncertainty for the the antineutrino flux  and an uncertainty of $0.5\%$~\cite{Baldoncini:2014vda} in the relative contribution of each isotope to the reactor fuel. 

\begin{figure}[ht]
\includegraphics[width=0.65\textwidth]{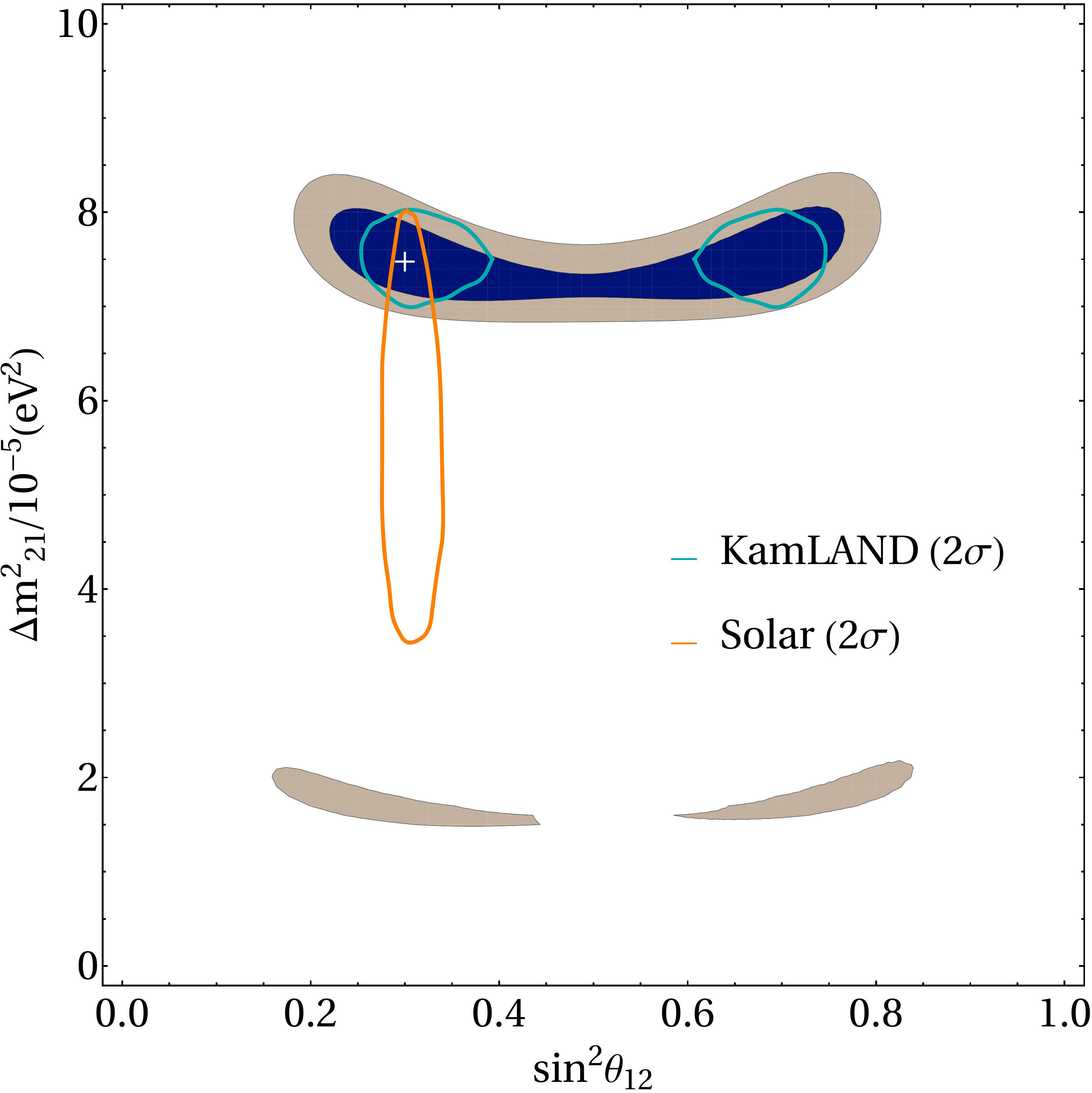}
\caption{Solid: 1 and the 2 sigma confidence regions of the $\sin^2\theta_{12}\times\Delta m^2_{21}$-parameter space allowed after analyzing one year of simulated reactor antineutrino data in SK. The input values for the solar parameters are $\Delta m^2_{21} = 7.53\times 10^{-5}\text{eV}^2$, $\sin^{2} \theta_{12}=0.30$, indicated with a plus-sign. The light blue  and orange contours are the regions allowed by KamLAND data~\cite{Gando:2013nba} and solar data~\cite{Abe:2016nxk}, respectively, at the two sigma level.}
\label{fig:xi2d}
\end{figure}

Fig.~\ref{fig:xi2d} depicts the region of the $\sin^2\theta_{12}\times\Delta m^2_{21}$-parameter space allowed after analyzing one year of simulated reactor antineutrino data in SK. The input values for $\sin^2\theta_{12}\times\Delta m^2_{21}$ are the best-fit values reported by KamLAND~\cite{Gando:2013nba}: $\Delta m^2_{21} =
7.53\times 10^{-5}\text{eV}^2$, $\sin^{2} \theta_{12}=0.30$. $\sin^2\theta_{13}$ has been fixed to $\sin^2\theta_{13} = 0.0224$, the current best-fit value according to Ref.~\cite{Esteban:2018azc}. 

One year of SK reactor data are insufficient to restrict $\sin^2\theta_{12}\times\Delta m^2_{21}$ to one continuous region; instead two isolated ``islands'' are allowed. The reason is the lack of statistics and the relatively poor energy resolution. Nonetheless, SK has good sensitivity to the value of $\Delta m^2_{21}$. 

\begin{figure}[ht]
\includegraphics[width=0.45\textwidth]{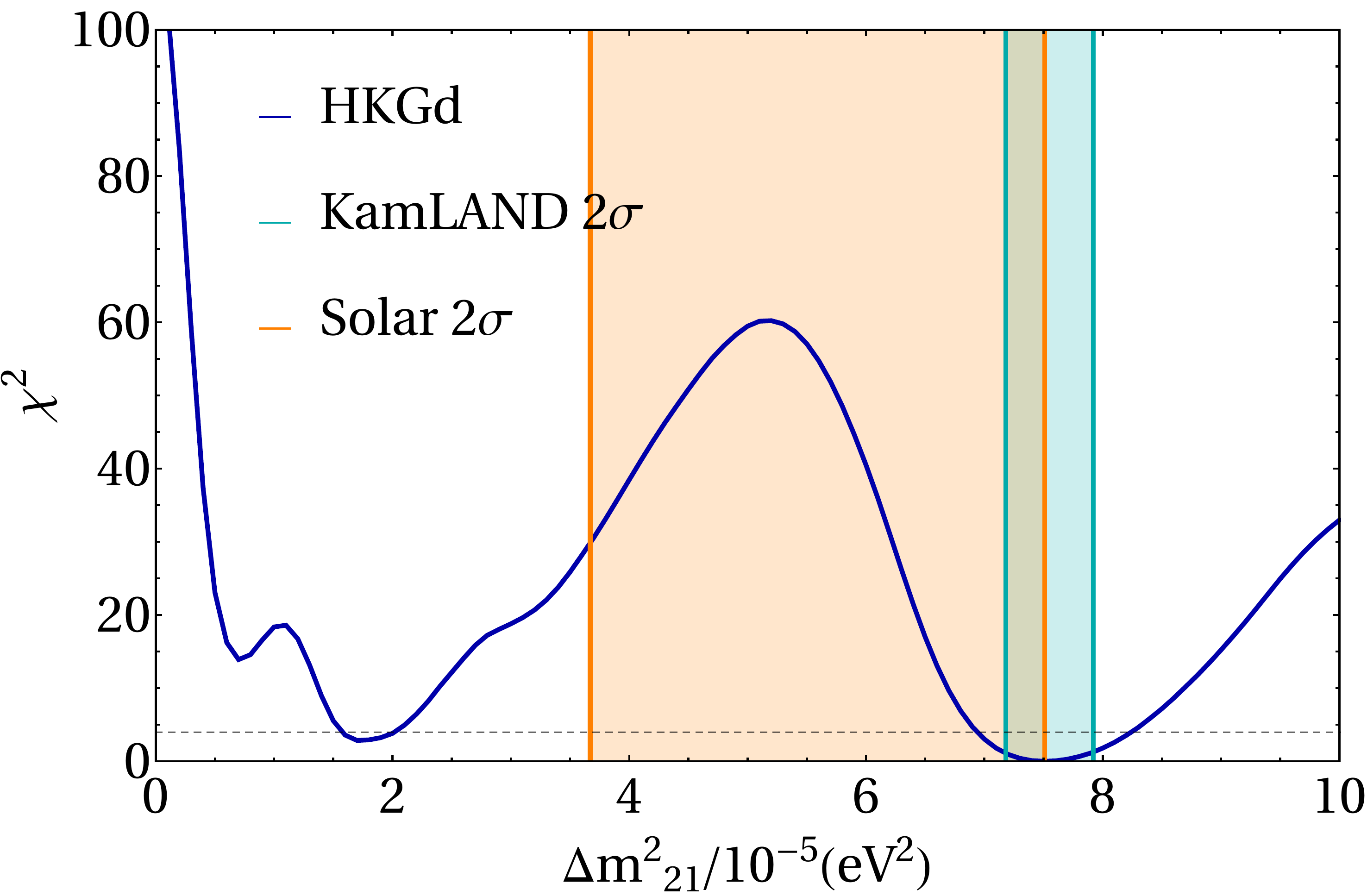}\,\includegraphics[width=0.45\textwidth]{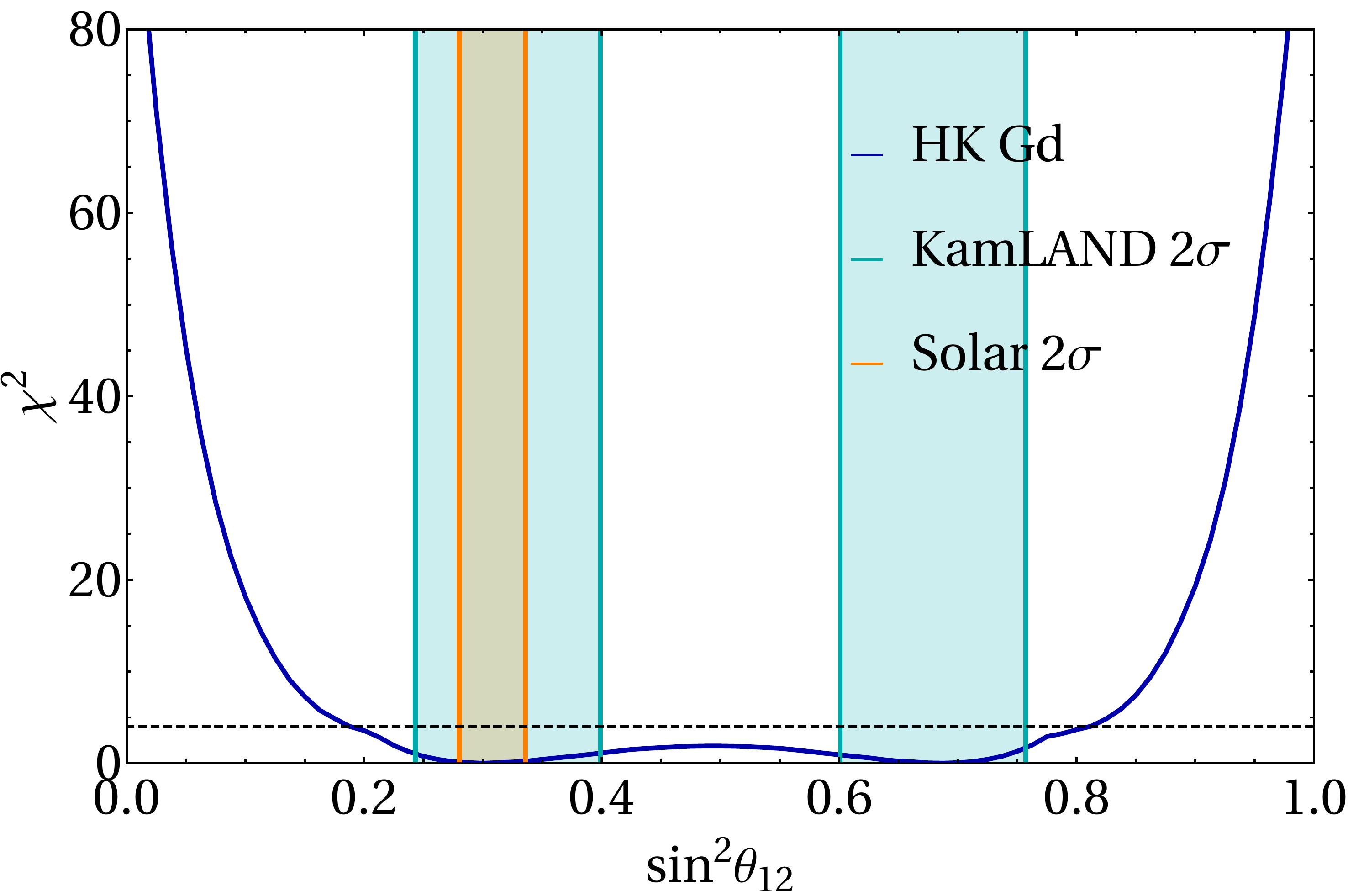}
\caption{$\chi^2$, relative to its minimum, as a function of $\Delta m^2_{21}$ marginalized over $\sin^2\theta_{12}$ (left) and
  $\sin^22\theta_{12}$ marginalized over $\Delta m^2_{21}$ (right), obtained from one year of simulated reactor antineutrino data in SK. The input values for the solar parameters are $\Delta m^2_{21} = 7.53\times 10^{-5}\text{eV}^2$, $\sin^{2} \theta_{12}=0.30$.}
\label{fig:xi1d}
\end{figure}

Fig.~\ref{fig:xi1d}(left) depicts the reduced $\chi^2$ as a function $\Delta m^2_{21}$, relative to its minimum value, marginalized over $\sin^2\theta_{12}$ while Fig.~\ref{fig:xi1d}(right) depicts the reduced $\chi^2$ as a function $\sin^2\theta_{12}$, relative to its minimum value, marginalized over $\Delta m^2_{21}$. The figure reveals that SK can establish that there are oscillations, and that the mixing angle is large: $\sin^2\theta_{12}\in [0.2,0.8]$, roughly, at the two-sigma level. The ``oscillations'' in Fig.~\ref{fig:xi1d}(left) indicate that degenerate values of $\Delta m^2_{21}$ can fit the data reasonably well. Nonetheless, if data at SK are consistent with the best-fit value preferred by KamLAND data, the best-fit value currently preferred by solar neutrino data -- $(\Delta m^2_{21})_{\rm solar}\sim 5\times 10^{-5}\text{eV}^2$ -- is ruled out at more than the seven-sigma level. 
 
In Fig.~\ref{fig:xi1d}, the parameter values highlighted in blue and orange are allowed, at the two-sigma level, by KamLAND and solar data, respectively. 
After one year of SK reactor-antineutrino data, both the SK and KamLAND uncertainties are comparable, especially if one is interested in $\Delta m^2_{21}$. More broadly, reactor data at SK are enough to resolve the presence and position of one oscillation maximum. Hence some values of $\Delta m^2_{21}$ are allowed while others are safely excluded. The ability to detect prompt energies below 4.5~MeV~\cite{Simpson:2019xwo} is key when it comes to resolving between the best-fit of KamLAND and solar data. If the true value of $\Delta m^2_{21}$ coincides with the best-fit to the solar data,  $\Delta m^2_{21}=5\times 10^{-5}\text{eV}^{2}$, the energy distribution of the events peaks below 4~MeV, see Fig.~\ref{fig:event}. If SK is not able to identify events with prompt energies below this threshold, the shape distortion introduced by the oscillations can be compensated by the systematic uncertainties in the flux. 

\begin{figure}[ht]
\includegraphics[width=0.45\textwidth]{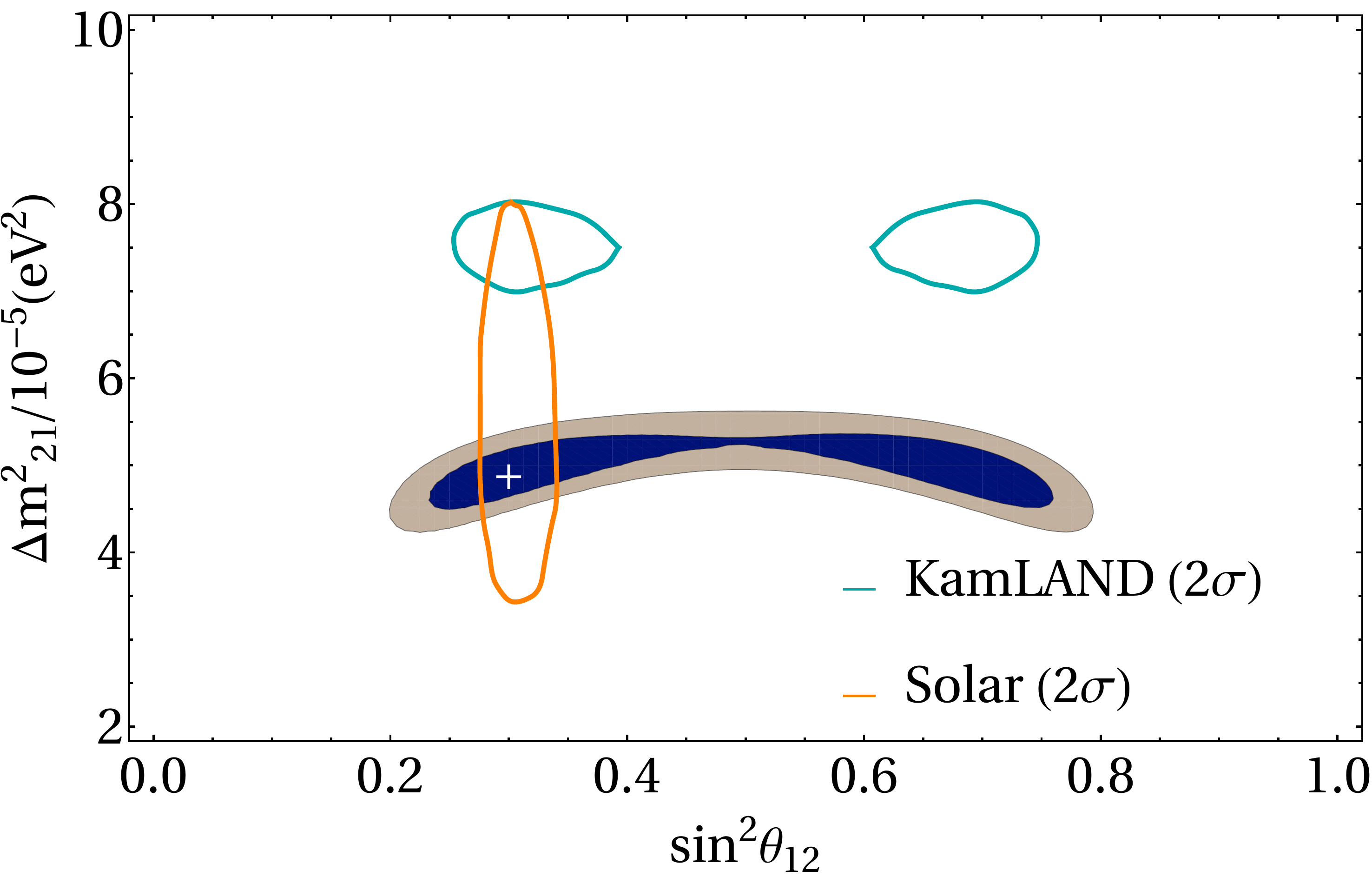}\,\includegraphics[width=0.45\textwidth]{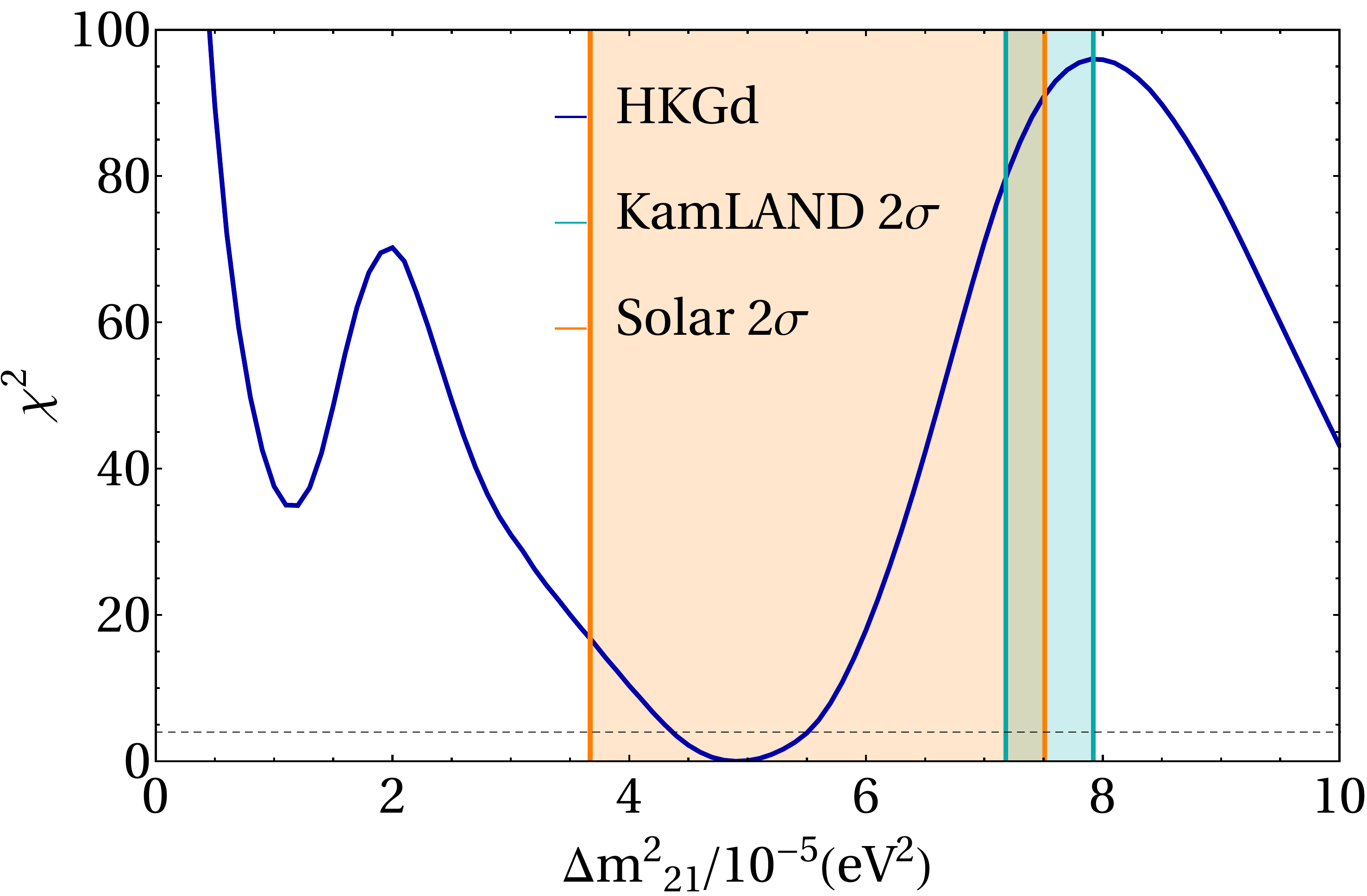}
\caption{Similar to Fig.~\ref{fig:xi2d}, on the left, and Fig.~\ref{fig:xi1d}(left), on the right. Here the input values of the solar parameters are $\Delta m^2_{21}=4.9\times 10^{-5}~\text{eV}^2$ and $\sin^2\theta_{21}=0.30$, indicated with a plus-sign on the left-hand panel.}
\label{fig:xisun}
\end{figure}

Under the same assumptions discussed above, Fig.~\ref{fig:xisun}(left) depicts the region of the $\sin^2\theta_{12}\times\Delta m^2_{21}$-parameter space allowed after analyzing one year of simulated reactor antineutrino data in SK when the input values for $\sin^2\theta_{12}\times\Delta m^2_{21}$ are the best-fit values extracted from solar neutrino data: $\Delta m^2_{21}=4.9\times 10^{-5}~\text{eV}^2$ and $\sin^2\theta_{21}=0.30$. $\sin^2\theta_{13}$ has been fixed to $\sin^2\theta_{13} = 0.0224$, the current best-fit value according to Ref.~\cite{Esteban:2018azc}. If the true value of $\Delta m^2_{21}$ coincided with the best-fit to the solar data, the best-fit value preferred by KamLAND would be ruled out at the nine sigma confidence level with one year of simulated SK reactor data. Fig.~\ref{fig:xisun}(right) depicts the reduced $\chi^2$ as a function $\Delta m^2_{21}$, relative to its minimum value, marginalized over $\sin^2\theta_{12}$. If SK reactor data were consistent with solar data, they would clash dramatically -- more than five sigma discrepancy -- with KamLAND results. 

After one year, we anticipate that SK will collect around $400$ reactor-antineutrino events, as depicted in Fig.~\ref{fig:event}. More statistics will lead to more precise results, improving on the estimates presented above. In the second half of the coming decade, however, a larger water Cherenkov detector will be able to measure the reactor antineutrino flux around the Kamioka mine, HyperKamiokande (HK). HK would contain $187$-kton of ultra-pure water and a photon-detector coverage similar to that of SK~\cite{Abe:2018uyc}, and there are plans to also dissolve Gd in the water. There is also the possibility that a second equivalent tank will be built in Korea (HKK). While the proposed location of HK is 8~km from the Kamioka site, the reactor neutrino flux there is, for all practical purposes, the same as the one observed in SK. For HKK, we assume it would be located in Mt. Bisul, in South Korea~\cite{Abe:2016ero}. We simulate the ability of HK and HKK to observe the nearby reactor antineutrino flux and measure the solar parameters. Our simulations are similar to the ones discussed for SK, with a few relevant differences. HK is planned to sit in a shallower site, hence the cosmic-ray-induced backgrounds are larger. Based on HK estimates~\cite{Abe:2018uyc}, we assume this background to be five times larger than what we assumed for SK, and we make the same assumption for HKK. After one year of running, assuming  the reactors that were running in 2018 will be running in the late 2020s, the number of expected events is around 3,200 for HK and 5,500 for HKK. With these data, we expect that HK can measure  $\Delta m^2_{21}$ with roughly $2\%$ precision while HKK would approach $0.9\%$ precision. Both measurements would supersede the sensitivity of KamLAND. For $\sin^2\theta_{12}$, the estimated precision is around $10\%$ and $ 3\%$ for HK and HKK, respectively. The estimated HKK precision supersedes that of the current solar data. Of course, by the time HK or HKK take data, it is expected that JUNO will have produced results that are at least as precise as the estimates summarize here.

In summary, given the size of SK and the number of nuclear reactors in its vicinity, we argued that SK doped with gadolinium can observe the oscillations of reactor antineutrinos driven by $\Delta m^2_{21}$. After only one year of data taking, we estimated that SK can measure $\Delta m^2_{21}$ with enough precision to help inform the current small tension between existing results from KamLAND and solar neutrino experiments. Our estimates are depicted in Figs.~\ref{fig:xi2d} and \ref{fig:xi1d}, for simulated data consistent with KamLAND's best-fit point, and Fig.~\ref{fig:xisun}, for simulated data consistent with the best-fit point preferred by solar data.  These results could be available before JUNO takes data.

\section*{Acknowledgements}

The work of AdG is supported in part by the DOE Office of Science
award \#DE-SC0010143. Fermilab is operated by the Fermi Research
Alliance, LLC under contract No. DE-AC02-07CH11359 with the United
States Department of Energy. IMS acknowledge travel support from the
Colegio de Fisica Fundamental e Interdisciplinaria de las Americas
(COFI).

\bibliography{biblio}

\end{document}